\newcommand{\CLASSINPUTtoptextmargin}{1in}
\documentclass[10pt,conference,final,a4paper]{IEEEtran}

\usepackage{times,amsmath}
\usepackage{cite}
\usepackage{float}
\usepackage{amsfonts}
\usepackage{amsmath} 
\usepackage{multirow}

\usepackage{array}
\usepackage{tikz}
\usepackage{graphicx}
\usepackage{caption}
\usepackage{subcaption}
\usepackage{siunitx}
\usepackage{booktabs}
\usepackage{epstopdf}
\usetikzlibrary{arrows,automata}

\makeatletter
\g@addto@macro\normalsize{%
  \setlength\abovedisplayskip{5pt}
  \setlength\belowdisplayskip{5pt}
  \setlength\abovedisplayshortskip{5pt}
  \setlength\belowdisplayshortskip{5pt}
}

\usepackage[caption=false,font=footnotesize]{subfig}
\usepackage{algpseudocode}
\usepackage{algorithm}

\usepackage{tikz}
\usetikzlibrary{arrows}

\usepackage{enumerate}

\title{Sleep Period Optimization Model For Layered Video Service Delivery Over eMBMS Networks}

\author{\IEEEauthorblockN{Lorenzo Carl\`a\IEEEauthorrefmark{1}, Francesco Chiti\IEEEauthorrefmark{1}, Romano Fantacci\IEEEauthorrefmark{1}, Andrea Tassi\IEEEauthorrefmark{3}
\IEEEauthorblockA{\IEEEauthorrefmark{1}Department of Information Engineering, University of Florence, Firenze, Italy}}
\IEEEauthorblockA{\IEEEauthorrefmark{3}School of Computing and Communications, Lancaster University, Lancaster, UK}}

\begin{document}
\maketitle
\begin{abstract} 
Long Term Evolution-Advanced (LTE-A) and the evolved Multimedia Broadcast Multicast System (eMBMS) are the most promising technologies for the delivery of highly bandwidth demanding applications. In this paper we propose a green resource allocation strategy for the delivery of layered video streams to users with different propagation conditions. The goal of the proposed model is to minimize the user energy consumption. That goal is achieved by minimizing the time required by each user to receive the broadcast data via an efficient power transmission allocation model. A key point in our system model is that the reliability of layered video communications is ensured by means of the Random Linear Network Coding (RLNC) approach.  Analytical results show that the proposed resource allocation model ensures the desired quality of service constraints, while the user energy footprint is significantly reduced.
\end{abstract}

\section{Introduction}
\label{Intro}
The diffusion of multimedia capable devices, such as smartphones and tablets, has generated a rapid growth of multimedia service demand. In particular, by 2018 the video traffic will represent $69\%$ of the mobile Internet data traffic.
This surge in demand has been addressed in fourth generation (4G) LTE-Advanced (LTE-A) cellular networks through the adoption of the evolved Multimedia Broadcast Multicast Service (eMBMS) framework~\cite{sesia}. In particular, a broadcast/multicast multimedia content can be delivered to the users as an eMBMS traffic flow in two modes: the Single Cell-eMBMS (SC-eMBMS) mode or the Single Frequency Network-eMBMS (SFN-eMBMS) mode~\cite{sesia}. In the former mode, each base station independently selects the communication parameters for the transmission of the multimedia content, such as, the amount of radio resources to be allocated, the power transmission, etc. Conversely, in the SFN-eMBMS mode, two or more space contiguous base stations, forming the SFN, transmit the same eMBMS flow in a synchronous fashion. Since base station of the same SFN do not interfere with each other, both the user Signal-to-Interference plus Noise Ratio (SINR) and the system spectral efficiency are significantly improved compared to the SC-eMBMS mode.

Among the issues related to the multicast/broadcast service delivery, the amount of energy required by a user to receive an eMBMS flow is of paramount importance~\cite{6461191}. For instance, during the reception of high data rate video streams, the radio interface of a user has to be in an active state for a time interval (hereafter referred to as \textit{activity period}) which could not be negligible. Furthermore, as the activity period increases, the energy footprint on the user battery increases as well~\cite{6151867}. In order to mitigate that issue in the case of Point-to-Point (PtP) communications, the serving base station~\cite{6461191,5373741}: (i) concentrates the data transmissions in short time intervals and, (ii) signals to the user to turn off its radio interface for a predetermined period of time, called \emph{sleep period}. It is straightforward to note that longer sleep periods can help to reduce the user battery consumption but it also increases the time interval when the user cannot receive any data from the network. For instance, J.-M. Liang~\textit{et al.}~\cite{6472116} attempts to maximise the sleep period duration of users involved in PtP communications. That optimization provides a generic trade-off between activity and sleep period duration suitable for a set of independent flows which may be received by each user. On the other hand, C. Wu~\textit{et al.}~\cite{6154772} propose to reduce the user energy footprint by decreasing the transition times between the sleep period and active period. The aforementioned proposal refers to a system model, where both PtP and Point-to-Multipoint (PtM) data flows are conveyed to the users. We remark that the optimization model proposed in~\cite{6154772} refers to best effort non-multimedia data flows. Unfortunately, a little attention has been paid to the problem of reducing the user energy consumption during the reception of multimedia eMBMS flows.

This paper focuses on the broadcasting of layered video services over SFN-eMBMS networks. In particular, we refer to video streams encoded by using the H.264 Scalable Video Coding compression standard (H.264/SVC)~\cite{6025326}. A H.264/SVC video service consists of one \textit{base layer} that provides a basic reconstruction quality and one or more \textit{enhancement layers}. Each user can improve the quality of the recovered video stream by combining the basic layer with one or more enhancement layers. In particular, we propose a green resource allocation framework aiming at minimizing the transmission time duration of a layered video stream and, hence, the user energy footprint. That goal is achieved by optimizing the power transmission associated to each video layer. Intuitively, if the power transmission associated to a video layer increases, then the video service can be delivered in a shorter time interval because of the adoption of a higher modulation order, lower coding rate, etc. As a consequence, the user sleep period are expected to increase and the user energy consumption decreases.

A key point of our system model is that reliability of PtM communication is ensured by the adoption of the Random Linear Network Coding (RLNC) approach~\cite{jsacTassi}. Usually, the eMBMS framework adopts Application Layer-Forward Error Correction (AL-FEC) codes based on Raptor codes~\cite{6353684}. However, this kind of codes are usually applied over large source messages. Hence, the adoption of Raptor or traditional LT codes could lead to a significant communication delay, as noted in~\cite{6416071}. That issue can be overcome by using the RLNC over short source messages~\cite{6353397}. In this paper, the proposed optimization framework ensures that predetermined fractions of users can recover the desired set of video layers with a target probability. In addition, the user energy efficiency goal of our model is achieved by jointly optimizing the transmission parameters and RLNC parameters.

The rest of the paper is organized as follows. Sec.~\ref{sys_model} describes the considered system model. Sec.~\ref{optimization_prob} defines the proposed optimization framework. In addition, Sec.~\ref{optimization_prob} provides also an efficient heuristic strategy that can derive a good quality feasible solution to the proposed optimization framework in a finite number of steps. Analytical results are provided in Sec.~\ref{Results}. Finally, we draw our conclusion in Sec.~\ref{Conc}.

\section{System Model}
\label{sys_model}
We consider a network scenario composed by $\Hat{B}$ contiguous base stations, where  $B_{\text{SFN}}$ out of $\Hat{B}$ base stations form the SFN area. In addition, in the rest of the paper, we assume that a H.264/SVC video stream is transmitted to a group of users $U$ over the SFN. According to the H.264/SVC compression standard, the video stream can be modelled as a set of Group of Pictures (GoPs), each of which has a fixed time duration of $t_{GoP}$ seconds~\cite{6025326}. The video stream is composed by $L$ layers $\{v_1, \ldots, v_L\}$, where $v_1$ is the base layer and $v_\ell$ (for $\ell = 2, \ldots, L$) the $\ell$-th enhancement layer. In addition, throughout the paper, we define the user Quality of Service (QoS) as the number of consecutive video layers that can be successfully recovered by a user, starting from the base layer.

\subsection{Layered Video Transmission over eMBMS Network}
\begin{figure}[tbd] 
\begin{center}
\includegraphics[width=0.75\columnwidth]{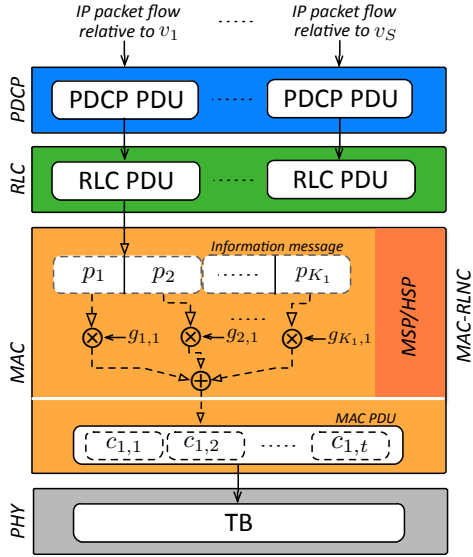}
\caption{The considered LTE-A protocol stack.}
\label{fig.stackNG}
\end{center}
\vspace{-5mm}
\end{figure}

This paper refers to the LTE-A system design proposed in~\cite{6353397} and sketched in Fig.~\ref{fig.stackNG}, where the reliability of the service delivery is improved by the adoption of the RLNC. In particular, all the RLNC related operations are performed by a coding sublayer, called MAC-RLNC, placed at the top of the standard LTE-A Medium Access Control (MAC) layer. 

We assume that each video layer is associated to an independent IP packet stream. All the $L$ layers that compose the video stream enter the Packet Data Conversion Protocol (PDCP) layer which produces a PDCP Protocol Data Units (PDCP PDU) stream. Then the Radio Link Control (RLC) layer segments/concatenates each PDCP PDU in order to produce a stream of RLC PDUs per video layer, which is forwarded to the MAC layer. According to the RLNC principle, the MAC-RLNC layer segments the $\ell$-th layer of each GoP into a stream of  $K_\ell$ \textit{information elements} $\{x_1, \ldots, x_{K_\ell}\}$. Then, all the information elements defining the video layer $\ell$ are linearly combined by the MAC-RLNC in order to generate a stream of $N_\ell \geq K_\ell $ \textit{coded elements}. The $j$-th coded element associated with the $\ell$-th video layer of a GoP is defined as $ c_{\ell,j} \! = \! \sum_{j=1}^{K_\ell} g_{\ell,j} \! \cdot \! x_j $, where $ g_{\ell,j} $ is a \textit{coding coefficient} uniformly selected at random over a finite field of size $q$~\cite{fragouli}. The stream of coded elements are mapped onto one or more MAC PDUs, which are delivered to the Physical layer (PHY) for the transmission. The layer $\ell$ of a GoP is recovered as soon as a user recovers $K_\ell$ linearly independent coded elements.

\begin{figure}[tbd]
\begin{center}
\includegraphics[width=0.77\columnwidth]{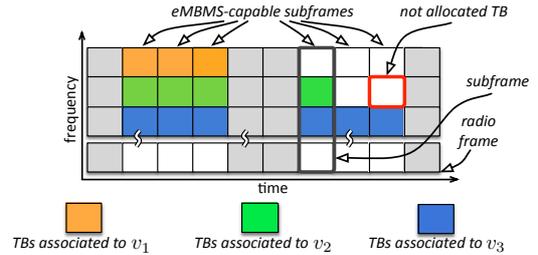}
\caption{LTE-A radio frame model and the considered eMBMS flow allocation.}
\label{fig.radioFrame}
\end{center}
\vspace{-5mm}
\end{figure}

The frequency-structure of a LTE radio frame consists of 10 subframes, where the time duration of each subframe is one TTI, namely, $t_{\emph{TTI}} = 1 \text{ms}$. In addition, as shown in Fig.~\ref{fig.radioFrame}, the LTE-A standard imposes that at most 6 out of 10 subframe per radio frame can be used to deliver eMBMS flows. In particular, we assume that the ratio of eMBMS-capable subframes per radio frame is $T_{\emph{MBMS}} = 0.6$. The basic data units delivered to a user in each subframe are called Transport Blocks (TBs). Each TB is transmitted by using a specific Modulation and Coding Scheme (MCS) and consists of one or more Resource Block Pairs (RBPs)~\cite{sesia}. In this paper, we assume that: (i) each TB is composed by the same number of RBPs, (ii) a TB can hold data associated to one video layer and, (iii) at most one TB associated with the same video layer can be mapped onto the same eMBMS-capable subframe (see Fig.~\ref{fig.radioFrame}).  
In our analysis, we assume that the user radio interface is active until each layer of a GoP is delivered. Hence, we define the user sleep period $\xi$ as follows:
\begin{equation}
\xi = d_{\emph{GoP}} - \max_{\ell = 1, \ldots, L}(t_\ell)
\end{equation}
where $d_{\emph{GoP}} = \frac{t_{\emph{GoP}} \, \cdot \, T_{\emph{MBMS}}}{t_{\emph{TTI}}}$ is the time duration of a GoP expressed in terms of number of TTIs; while $t_\ell$ is the number of TB transmissions associated to the layer $\ell$ of a GoP.

%The number of TBs allocated and the power transmission used for the delivery of the $\ell$-th layer of a GoP are denoted by $t_\ell$ and $P_\ell$, respectively.

%In particular, the pair ($P_\ell$, $t_\ell$) defines the \textit{radio resource footprint} associated to the transmission of the $\ell$-th layer of a GoP. Furthermore, from Fig.~\ref{fig.radioFrame}, the $L$ layers that compose the video service may have different values of $t_\ell$ (according to the QoS requirements of each layer). 

\subsection{Performance Evaluation} 
In order to efficiently estimate the QoS level experienced by a user, it is useful to model the data rate associated with the reception of a TB stream by means of the Shannon's rate formula:   
\begin{equation}
r(\sigma) = \alpha \, W \, log_2(1 + \beta \, \sigma)
\label{eq_shannon}
\end{equation}
where $W$ is the bandwidth spanned by a TB and $\sigma$ is the SINR associated with the considered user. The terms $\alpha$ and $\beta$ are non-negative correction factors which adapt the Shannon rate expression to fit to the actual user reception rate of a TB stream. In particular, parameters $\alpha$ and $\beta$ can be found by solving the mathematical regression problem presented in the Appendix.

In a SFN-eMBMS network, the $B_{\text{SFN}}$ base stations belonging to the SFN are coordinated in order to deliver the same physical signal. Hence, the signals coming from all the base stations of the SFN area are treated by a user as well as multipath components of a single base station transmission~\cite{sesia}. On the other hand, the remaining $\Hat{B} - B_{\text{SFN}}$ base stations interfere with the user reception. 
The channel gain experienced by user $u$ that is receiving from base station $i$ is denoted by $h_{u,i}$ and can be defined as follows\cite{6353397}: 
\begin{equation}
h_{u,i} = G_{\text{Tx}} + G_{\text{Rx}} - \text{PL}_{u,i} - \delta_{i} \, I_{u,i} 
\label{eq_chan_gain}
\end{equation}
where $G_{\text{Tx}}$ and $G_{\text{Rx}}$ are the antenna gains at the transmitter and receiver side, respectively. The term $\text{PL}_{u,i}$ is the path-loss experienced by user $u$ and associated with base station $i$ (see \mbox{Table \ref{tab.sim_parameters}}), while $N$ is the noise power. The terms $I_{u,i}$ denotes the Inter-Cell Interference and $\delta_i \in \lbrace 0 , 1 \rbrace$ is a binary parameter indicating if the $i$-th base station belongs to the SFN or not. In particular, we have that $\delta_i = 1$ if base station $i$ belongs to the SFN, otherwise $\delta_i = 0$. For these reasons, the SINR value experienced by a user receiving the $\ell$-th video layer can be expressed as follows:
\begin{align}
\sigma_{u,\ell} = \overbrace{\left(\frac{\displaystyle\sum_{i=1}^{B_{\text{SFN}}} h_{u,i}}{\displaystyle\sum_{j=1}^{\Hat{B} - B_{\text{SFN}}} h_{u,j} \,\, P_{t} +N}\right)}^{H_{u}} \,\, P_\ell = H_{u} \,\, P_\ell 
\label{eq_sinr}
\end{align}
where $P_\ell$ is the transmission power used to broadcast each TB related to the $\ell$-th layer of a GoP. The terms $h_{u,i}$ and $h_{u,j}$ denote the channel gains associated with the $i$-th base station belonging to the SFN and the $j$-th interfering base station, respectively. Finally, $P_t$ is the transmission power of any interfering base station.

From~\eqref{eq_shannon},~\eqref{eq_sinr} and the regression model presented in the Appendix, the rate experienced by the $u$-th user during the reception of a TB related to the $\ell$-th layer can be restated as a function of $P_\ell$:
\begin{equation} 
%\hspace{-0.2mm}
r_u(P_\ell) \!\! = \!\! \begin{cases}
			0\text{,} & \!\!\! \text{if $H_{u} P_\ell \! < \! \sigma_{\text{min}}$}\\
			\alpha  W  \log_2(1+ \beta  H_{u}  P_\ell )\text{,} & \!\!\! \text{if $\sigma_{\text{min}} \! \leq \! H_{u} P_\ell \! \leq \! \sigma_{\text{max}}$}\\
			\alpha W  \log_2(1+ \beta  \sigma_{\text{max}} )\text{,} & \!\!\! \text{if $H_{u}P_\ell \! > \! \sigma_{\text{max}}$.}
		 \end{cases}\!\!
\label{eq_shannon_correct_2}
\end{equation}
For this reason, from~\eqref{eq_shannon_correct_2} and according to the network coded eMBMS service delivery approach, the average number of coded elements associated to the $\ell$-th layer of a GoP that are received by user $u$ is 
\begin{align}
N_\ell(P_\ell,t_\ell) = \Bigg\lfloor\frac{t_{\emph{TTI}} \,\, r_u(P_\ell) \,\, t_\ell}{L_\ell}\Bigg\rfloor
\label{eq.recvSym}
\end{align}
where $L_\ell$ is the bit size of a coded element.
Furthermore, the probability that user $u$ is able to recover the $\ell$-th layer of a GoP, i.e., the probability that user $u$ is able to collect $K_\ell$ linearly independent coded elements out of $N_\ell$, can be expressed as follows~\cite{5634159}:
\begin{align}
g_u(P_\ell,t_\ell)=\displaystyle\prod_{j = 0}^{K_\ell-1} \left[ 1 - \frac{1}{q^{N_\ell(P_\ell,t_\ell)-j}} \right]. 
\label{probReception}
\end{align}  
In the following section, we will show how the pair \mbox{($P_\ell$, $t_\ell$)} can be optimized in order to ensure that a predetermined fraction of users can achieve the QoS level $\ell$ with at least a target probability. 

\section{Sleep Period Resource Allocation Framework}
\label{optimization_prob}
In this section we propose an energy-efficient radio resource allocation strategy called ``Maximum Sleep Period'' (MSP) that aims at maximizing the user sleep period by minimizing the number of TB transmissions needed for the transmission of a GoP. In addition, the proposed model ensures that a predefined fraction of users can recover the first $\ell$ layers with a given probability, for $\ell = 1, \ldots, L$.
To this end, for each video layer the MSP model (i) minimizes the maximum number of TB transmissions per video layer and, (ii) selects the power transmission $P_\ell$, for for $\ell = 1, \ldots, L$, such that the overall transmission power is less than or equal to a threshold value $\Hat{P}$. Hence, the MSP model is defined as follows:
\begin{align} 
	\text{(MSP)} &  \,  \mathop{\mathop{\mathop{\min\max}_{\ell \in \{1, \ldots, L\}}}} \,\,  t_\ell \label{ra.of}\\
    \!\!\!\!\text{subject to}&   \, \sum_{u = 1}^U \Delta \Big( g_u(P_\ell, t_\ell) \! \geq \! \Hat{\Phi} \Big) \! \geq \! \Hat{\theta}_\ell U \,\,\,\,\, \text{$\ell \in \{1, \ldots, L\}$} \label{ra.c1}\\
    	& \, K_\ell \le t_\ell \leq d_{\emph{GoP}} \quad\quad\quad\quad\quad\,\,\,\,\,\,\,\,\, \text{$\ell \in \{1, \ldots, L\}$} \label{ra.c2}\\
    & \,\sum_{\ell=1}^L P_\ell \leq \Hat{P} \label{ra.c3}\\
    & \, P_\ell \in \mathbb{R}^+, t_\ell \in \mathbb{N} \label{ra.c4} \quad\quad\quad\quad\quad\,\,\,\,\,\, \text{$\ell \in \{1, \ldots, L\}$}
\end{align}
where $\Delta(s)$ is an indication function such that $\Delta(s) = 1$ if the statement $s$ is true, otherwise $\Delta(s) = 0$. The constraint~\eqref{ra.c1} ensures that the fraction of users recovering layer $\ell$ with, at least, a probability of $\Hat{\Phi}$ is at least equal to $\Hat{\theta}_\ell$. Constraint~\eqref{ra.c2} ensures that the number of TB transmissions associated with layer $\ell$ does not exceed the time duration of a GoP. Finally, constraint~\eqref{ra.c3} imposes that the overall transmission power does not exceed the threshold $\Hat{P}$. Unfortunately, the coupling constraint~\eqref{ra.c3} turns MSP into a computationally complex mixed integer non-linear optimization problem. In order to mitigate this issue, in the rest of the section, we propose the Heuristic-MSP (H-MSP) strategy that provides a good quality feasible solution to MSP, in a finite number of steps.

For the sake of our analysis, we define from MSP the Unconstrained Sleep Period (USP) problem by simply removing the constraint~\eqref{ra.c3}. The solution of the aforementioned problem can be easily found by decomposing the USP problem into $L$ independent subproblems, where the $\ell$ subproblems can be expressed as follows:
\begin{align}
	\text{(USP-$\ell$)} &  \quad\quad  \mathop{\mathop{\min}} \,\,  t_\ell \label{ra2.of}\\
    \text{subject to} &   \quad\quad \sum_{u = 1}^U \Delta \Big( g_u(P_\ell, t_\ell) \geq \Hat{\Phi} \Big) \! \geq \! \Hat{\theta}_\ell U \label{ra2.c1}\\
    	& \quad\quad K_\ell \le t_\ell \leq d_{\emph{GoP}} \label{ra2.c2}\\
    & \quad\quad P_\ell \in \mathbb{R}^+, t_\ell \in \mathbb{N} \label{ra2.c3}
\end{align}
From the analysis presented in~\cite{6550868}, we understand that the optimum solution of USP-$\ell$, if it exists, belongs to the set $\mathcal{L}_\ell \doteq \Big\{ (P_\ell, t_\ell) \in \mathbb{R}^{+} \times \mathbb{N}\,\, \Big| \,\, K_\ell \leq t_\ell \leq d_{GoP}  \,\, \wedge \,\, \sum_{u = 1}^U \Delta \Big( g_u(P_\ell, t_\ell) \Big) \geq \Hat{\Phi} ) \! \geq \! \Hat{\theta}_\ell U\Big\}$. Let $\mathcal{L}_\ell(t_\ell)$ be the transmission power value such that $(\mathcal{L}_\ell(t_\ell), t_\ell) \in \mathcal{L}_\ell$. Hence, the optimum solution of USP-$\ell$ is represented by the pair ($\mathcal{L}_\ell(\tilde{t}_\ell)$,$\tilde{t}_\ell$) which is characterized by the minimum value of $t_\ell$ among all the possible pairs in $\mathcal{L}_\ell$~\cite{6550868}. The optimal solution to the USP model is given by $\{(\mathcal{L}_1(\tilde{t}_1), \tilde{t}_1), \ldots, (\mathcal{L}_L(\tilde{t}_L), \tilde{t}_L)\}$, for $\ell = 1, \ldots, L$.

\setlength{\textfloatsep}{2pt}
\begin{algorithm}[tbd] 
\floatname{algorithm}{Procedure}
\caption{Heuristic Maximum Sleep Period Strategy.}
\label{proc.1}
{\footnotesize \begin{algorithmic}[1]
\State Initialize $P^{*}_\ell \gets \tilde{P}_\ell$ and $t^{*}_\ell \gets \tilde{t}_\ell$, for $\ell = 1, \ldots, L$ 

\While {$\displaystyle\sum_{\ell=1}^{L} P^{*}_\ell \geq \Hat{P}$}\label{proc.rWhile}

	\For {$\ell \gets 1, \ldots, L$} \label{proc.rFor}
		\If {$t^{*}_\ell + 1 \leq d_{GoP}$}
			\State $P^\prime_\ell \gets \mathcal{L}_\ell(t^{*}_\ell + 1)$
            \State $t^\prime_\ell \gets t^{*}_\ell + 1$
        \Else
        	\State $t^\prime_\ell \gets \infty$
        \EndIf
	\EndFor \label{proc.rEndFor}

	\If {$o^\prime_\ell = \infty$ } \label{proc.rIf10}	

    	\State \Return no solution \label{proc.rIfretRetNF}
    \EndIf \label{proc.rEndIf10}
    
    \State $i \gets \arg\min\{t^\prime_1, \ldots, t^\prime_L\}$ \label{proc.rArgMin}
    \State $t^{*}_i \gets t^{*}_i + 1$
    \State $P^{*}_i \gets P^\prime_i$ \label{proc.rup}
\EndWhile \label{proc.rEndWhile}

\State \Return $(P^{*}_\ell, t^{*}_\ell)$, for $\ell = 1, \ldots, L$ \label{proc.ret}

\end{algorithmic}}
\end{algorithm}

Because of the definition of the USP model, it is worth noting that its optimal solution could be infeasible from the point of view of the MSP problem, i.e. the constraint~\eqref{ra.c3} could not be met. However, starting from the USP solution, we define the H-MSP. The idea behind H-MSP is that of turning the optimal USP solution, which is infeasible from the point of view of MSP, into a feasible MSP solution. The aforementioned transformation is performed by altering one component of the optimal USP solution at a time. In particular, from~\eqref{probReception}, we have that $\mathcal{L}_\ell(\tilde{t}_\ell + 1) \leq \mathcal{L}_\ell(\tilde{t}_\ell)$. Hence, during each iteration, the H-MSP strategy increases the smallest value of $t_\ell$ among the video layers. As a consequence, the total transmission power is reduced after each iteration. The general definition of H-MSP is provided in Procedure~\ref{proc.1} and it involves the following steps:
\begin{enumerate}[(i)]
\item We set $\{(P^{*}_1, t^{*}_1), \ldots, (P^{*}_L, t^{*}_L)\}$ equal to the solution of the USP model. If the constraint \eqref{ra.c3} is met, then the procedure returns the solution of USP. In other words, the solution of the USP model is equal to the optimal solution of MSP. 
\item Otherwise, the while-loop body (lines \ref{proc.rWhile}-\ref{proc.rEndWhile} of Procedure~\ref{proc.1}) for each layer increases the value of $t^{*}_\ell$ and sets ${P^\prime_\ell = \mathcal{L}_\ell(t^{*}_\ell + 1)}$. Then, the solution component characterized by the smallest value of $t^\prime$ is altered (lines \ref{proc.rArgMin}-\ref{proc.rup}).
\item The while-loop body iterates until the constraint \eqref{ra.c3} is met.
\end{enumerate}
It is worth noting that the while-loop body (lines 2-17) returns a feasible solution of MSP in a number of steps which is less than or equal to $\sum_{\ell=1}^L (d_{\emph{GoP}} - K_{\ell}$).

\section{Numerical Results} \label{Results} 
In this section, we investigate the performance of the MSP and H-MSP models in terms of the normalized sleep period associated with the delivery of each GoP, defined as follows:
\begin{equation}
\epsilon \! = \! \frac{\xi}{d_{\emph{GoP}}} 
\end{equation}
In addition, we also evaluate the user performance in terms of the maximum Peak Signal-to-Noise Ratio (PSNR) $\overline{p}_{u}$ that user $u$ can achieve:
\begin{equation}
\overline{p}_{u} = \max_{\ell = 0,\ldots,L} { \Big\lbrace \Hat{p}_\ell \, g_u \Big\rbrace }
\end{equation}
where $\Hat{p}_\ell$ is the PSNR associated with the reconstruction of the first $\ell$ layers of the video service~\cite{6025326}.

\renewcommand{\arraystretch}{0.8}
\begin{table}[tbp]
\centering
\caption{Main simulation parameters.}
\label{tab.sim_parameters}
\begin{tabular}{|m{0.30\linewidth}|m{0.50\linewidth}|}
\hline
\multicolumn{1}{|c|}{\textbf{Parameter}} & \multicolumn{1}{l|}{\textbf{Value}}  \\ 
\hline
\multicolumn{1}{|c|}{Base Stations (SFN Area)}  & \multicolumn{1}{l|}{19 (4)} \\
\hline 
\multicolumn{1}{|c|}{Inter-Site Distance (ISD)} & \multicolumn{1}{l|}{$500$ m}   \\
\hline 
\multicolumn{1}{|c|}{System Bandwidth} & \multicolumn{1}{l|}{$20$ MHz} \\
\hline 
\multicolumn{1}{|c|}{Transmission Scheme} & \multicolumn{1}{l|}{SISO} \\
\hline 
\multicolumn{1}{|c|}{Antenna Gains} & \multicolumn{1}{l|}{See Table A.2.1.1.2-2 \cite{Rel9}} \\ 
\hline 
\multicolumn{1}{|c|}{Penetration Loss} & \multicolumn{1}{l|}{20 dB (See Table A.2.1.1.5-2 \cite{Rel9})} \\
\hline
\multicolumn{1}{|c|}{Path-Loss} & \multicolumn{1}{l|}{See Table A.2.1.1.5-2 \cite{Rel9}} \\
\hline
\multicolumn{1}{|c|}{Duplexing Mode} & \multicolumn{1}{l|}{FDD} \\
\hline
\multicolumn{1}{|c|}{Channel Model} & \multicolumn{1}{l|}{PedA} \\
\hline 
\multicolumn{1}{|c|}{Carrier Frequency} & \multicolumn{1}{l|}{$2.0$ GHz}   \\
\hline 
\multicolumn{1}{|c|}{Noise Power} & \multicolumn{1}{l|}{$-168 \text{ dBm/Hz}$}   \\
\hline
\multicolumn{1}{|c|}{Power Transmission}  & \multicolumn{1}{l|}{$20-80$ W  (Table A.2.1.1-2 \cite{Rel9})}  \\ \cline{2-2}
\hline 
\multicolumn{1}{|c|}{$\Hat{\Phi}$} & \multicolumn{1}{l|}{0.1} \\
\hline 
\multicolumn{1}{|c|}{$T_{\emph{MBMS}}$} & \multicolumn{1}{l|}{0.6} \\
\hline
\multicolumn{1}{|c|}{$q$} & \multicolumn{1}{l|}{$2^8$} \\
\hline
\multicolumn{1}{|c|}{Number of RBPs} & \multicolumn{1}{l|}{6, 9, 12} \\
\hline 
\multicolumn{1}{|c|}{$d_{\emph{GoP}}$} & \multicolumn{1}{l|}{$320$ TTI} \\
\hline
\multicolumn{1}{|c|}{\parbox[c]{9.5mm}{Video A \\ \,\,$\mbox{L} = 3$}} & \multicolumn{1}{l|}{\parbox[c]{49mm}{\vspace{1mm} $\Hat{p}_\ell \! = \! \lbrace 29.94, 34.78, 40.73 \rbrace [\mbox{dB}]$ \cite{Munaretto} \\ $\Hat{r}_\ell  = \lbrace 117.1, 402.5, 1506.3 \rbrace [\mbox{kbps}]$ \\ $\Hat{\theta}_\ell = \lbrace 0.3, 0.6, 0.9 \rbrace$}} \\
\hline 
\multicolumn{1}{|c|}{\parbox[c]{9.5mm}{Video B \\ \,\,$\mbox{L} = 4$}} & \multicolumn{1}{l|}{\parbox[c]{50.23mm}{\vspace{1mm} $\Hat{p}_\ell \!=\! \lbrace 29.45, 32.30, 34.52, 38.41 \rbrace [\mbox{dB}]$\cite{Munaretto} \\ $\Hat{r}_\ell = \lbrace 160.0, 300.0, 560.0, 1150.0 \rbrace [\mbox{kbps}] $ \\ $\Hat{\theta}_\ell = \lbrace 0.3, 0.5, 0.6, 0.9 \rbrace$}} \\
\hline
\end{tabular}
\vspace{3mm}
\end{table}

The proposed optimization framework is also compared against an Uniform Power Allocation (UPA) strategy. The considered UPA strategy transmits each video layer by using the same transmission power $P_\ell = \frac{\Hat{P}}{L}$, for $\ell = 1, \ldots, L$. In addition, for the sake of comparison, we assume that the UPA model relies on the MAC-RLNC coding sublayer. For these reasons, the UPA model can be defined as follows:
\begin{align}
	& \text{\hspace{-2mm}(UPA)} \quad\quad\quad\,\,\,  \mathop{\mathop{\min}} \,\,  t_\ell \label{upa.of}\\
    & \hspace{20mm} K_\ell \le t_\ell \leq d_{\emph{GoP}} \quad\quad\quad\,\,\,\, \text{$\ell \in \{1, \ldots, L\}$} \label{upa.c1}\\
    & \hspace{20mm} P_\ell \in \mathbb{R}^+, t_\ell \in \mathbb{N} \label{upa.c2} \quad\quad\quad\,\,\, \text{$\ell \in \{1, \ldots, L\}$}
\end{align}

\begin{figure}[tbd] 
\begin{center}
\includegraphics[width=0.88\columnwidth]{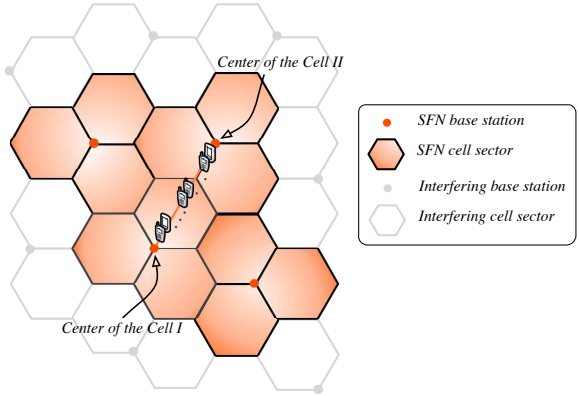}
\caption{A part of the considered SFN-eMBMS scenario.}
\label{fig.net_scenario}
\end{center}
\vspace{-3mm}
\end{figure}

Analytical results have been derived by considering a LTE-A network composed by $19$ base stations deployed in two concentric rings, with an inter-site distance of $500$ m. Each base station controls three hexagonal cell sectors. As shown in Fig.~\ref{fig.net_scenario}, we refers to a SFN composed by $4$ contiguous base stations, which is surrounded by the remaining $15$ interfering base stations. For what concerns the user distribution, we refer to a scenario characterized by a high heterogeneity in terms of propagation conditions. In particular, as reported in Fig.~\ref{fig.net_scenario}, we assume that users are placed on the symmetry axis of a sector of cell I of the SFN. In particular, we refer to $U = 80$ users, which are spaced $2$ m apart with the first user being placed $90$ m from the centre of cell I. Table~\ref{tab.sim_parameters} summarizes the main simulation parameters and the considered H.264/SVC video streams, namely video A and video B.

\begin{figure}[tbd] 
\begin{center}
\includegraphics[width=0.9\columnwidth]{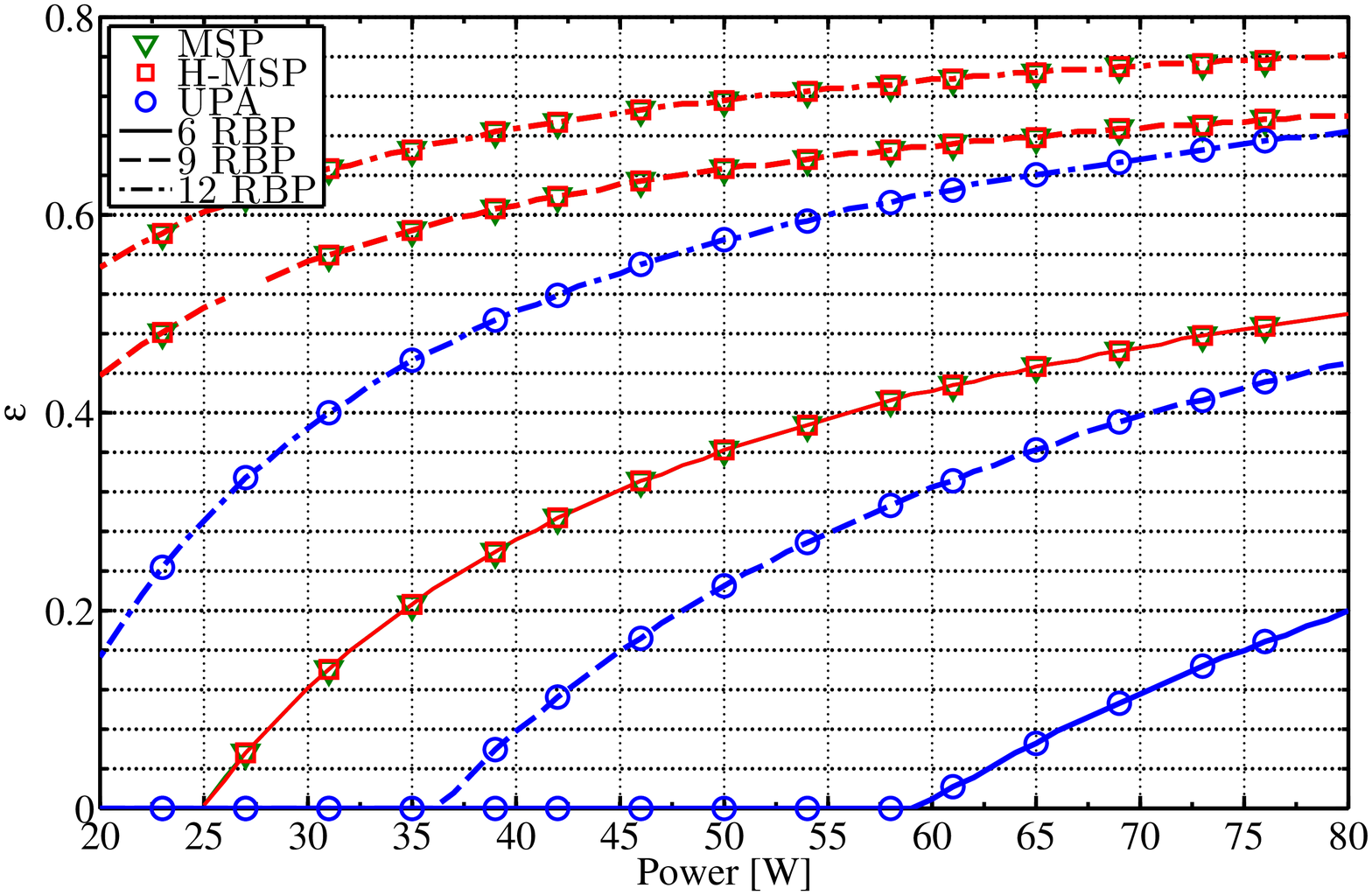}
\caption{Normalized sleep period duration vs. overall transmission power threshold, for video A.} 
\label{fig.cfrA}
\end{center}
\vspace{-1mm}
\end{figure}

\begin{figure}[tbd]
\begin{center}
\includegraphics[width=0.9\columnwidth]{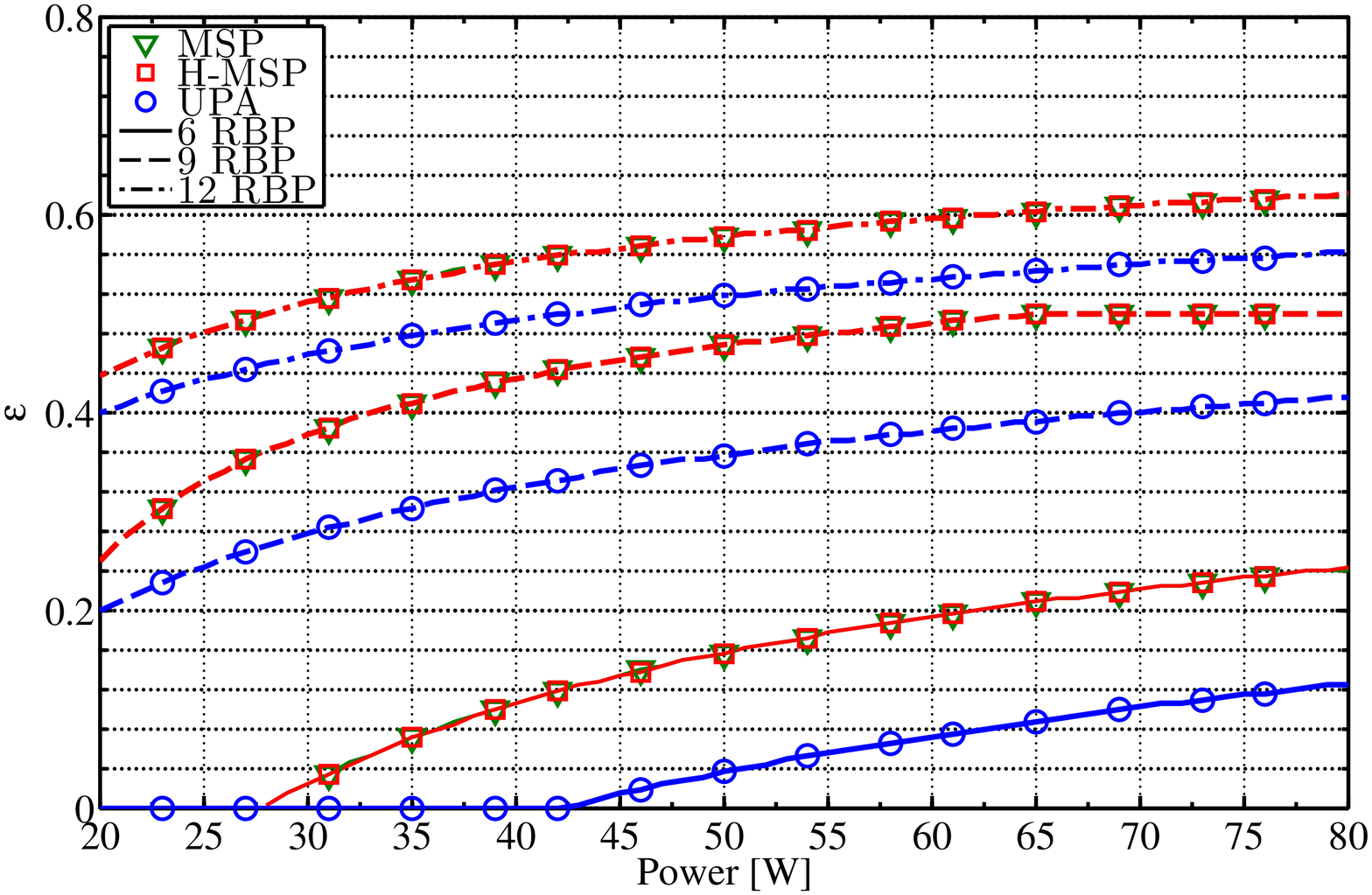}
\caption{Normalized sleep period duration vs. overall transmission power threshold, for video B.}
\label{fig.cfrB}
\end{center}
\vspace{-1mm}
\end{figure}

% \includegraphics[width=\columnwidth]{img/psnrA}
%                \caption{A gull}
%                \label{fig:gull}
%       
%                \includegraphics[width=\columnwidth]{img/psnrB.eps}
%                \caption{A tiger}
%                \label{fig:tiger}

\begin{figure}[t]
        \centering
        \begin{subfigure}[t]{\columnwidth}
        \includegraphics[width=\columnwidth]{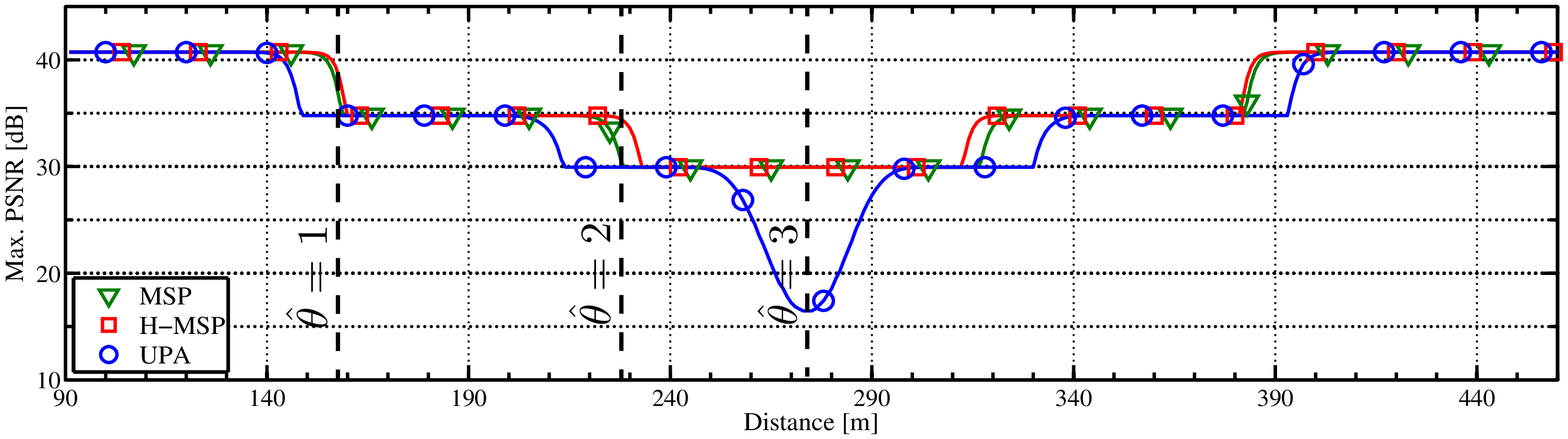}
        \vspace{-5mm}   	    
 	\caption{Video A}
 	\label{fig:psnrA}
		
	\vspace{3mm}
 	\includegraphics[width=\columnwidth]{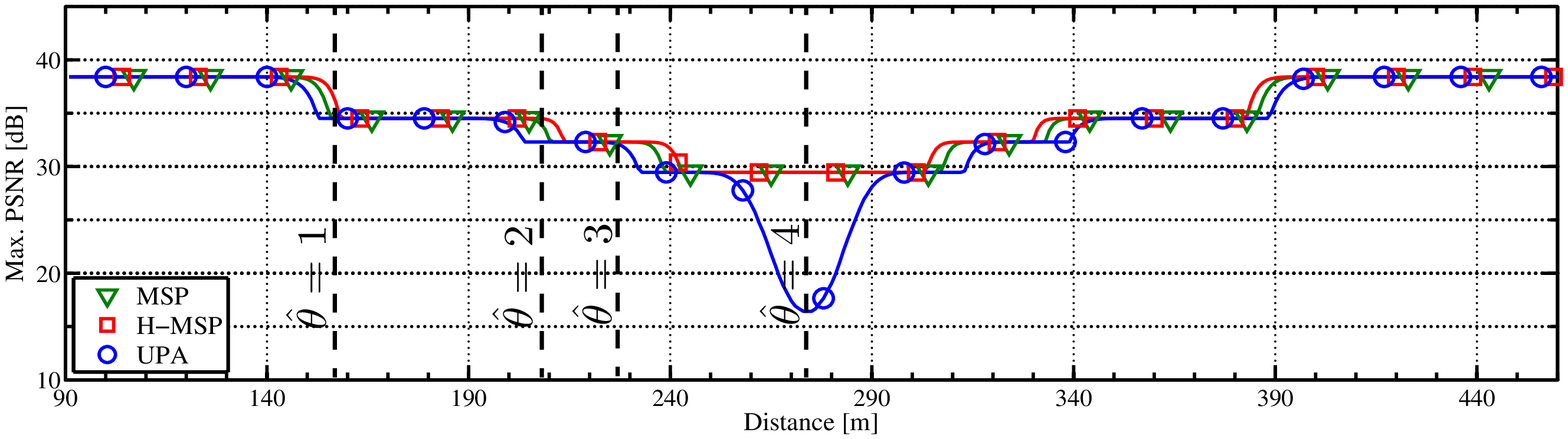}
        \caption{Video B}
        \label{fig:psnrB}
        \end{subfigure}
        \caption{Maximum achievable PSNR of video A and video B vs. distance from the centre of cell I, for $\Hat{P} = 40$ W.}

\label{fig:psnrAB}
\end{figure}

Fig.~\ref{fig.cfrA} compares performance of MSP, H-MSP and UPA strategies in terms of the normalized sleep period associated with the delivery of a GoP of video A as a function of the overall transmission power threshold $\Hat{P}$. In particular, we consider different TB sizes, namely, $6$, $9$, and $12$ RBPs. The effectiveness of the proposed H-MSP strategy is proved by the fact that the gap between the MSP and the H-MSP models is negligible. Furthermore, we note that as the value of $\Hat{P}$ and the number of RBPs increase, the H-MSP and MSP models always provide resource allocation solutions characterized by sleep periods that are up to 10\% greater than those associated with the UPA strategy. The normalized sleep period results relative to video B are provided by Fig.~\ref{fig.cfrB}. The figure shows a performance behaviour which is similar to what shown by Fig.~\ref{fig.cfrA}. In particular, the proposed MSP and H-MSP ensure sleep periods that are up to $40\%$ greater than those achieved by the UPA model. 

Fig.~\ref{fig:psnrAB} compares all the considered strategies in terms of the maximum user PSNR, in the case of TBs composed by $9$ RBPs and $\Hat{P} = 40$ W. Since users are regularly distributed and equally spaced among each other, the expression of $\overline{p}_{u}$ can be equivalently expressed as a function of the distance between user $u$ and the centre of cell I. For the same reason, parameters $\Hat{\theta}_1, \ldots, \Hat{\theta}_L$ can be equivalently interpreted as distances from the centre of cell I. In particular, the vertical dashed lines of Fig.~\ref{fig:psnrAB} mark the corresponding values of $\Hat{\theta}_\ell$, for $\ell = 1, \ldots, L$. For what concerns video A, Figure~\ref{fig:psnrA} shows that the UPA model delivers the base layer and all the video layers up to a distance (from the centre of cell I) which is $33$ m and $32$ m smaller than the distance provided by both the MSP and H-MSP strategies, respectively. On the other hand, Figure~\ref{fig:psnrB} shows the same performance metric in the case of video B. Also in this case, we understand that the service coverage provided by the UPA model is smaller than that of the developed strategies. 

Consider again Fig.~\ref{fig:psnrAB}, the maximum PSNR values reported for distances greater than $272$ m refer to a user (excluded from the optimization process) which moves on the extension of the considered cell-sector symmetry axis, towards the centre of cell II (see Fig.~\ref{fig.net_scenario}). It is worth noting that the resource allocation solutions provided by both the MSP and H-MSP strategies ensure (at least) the base layer of video A and B to be received while an user moves from cell I to cell II. On the other hand, neither in the case of video A nor video B the UPA strategy can achieve the same kind of service continuity.

\section{Conclusions}
\label{Conc}
In this paper we address the problem of minimizing the user energy consumption for video delivery over SFN-eMBMS networks. To this end, we propose an optimal and heuristic radio resource allocation strategy, namely MSP and H-MSP strategies, which maximize the user sleep period and improve the reliability of communications by means of an optimized RLNC approach. In this way, not only the the user energy consumption is reduced but also the developed strategies can meet the desired QoS levels.
Results show that the developed H-MSP strategy provide a good quality feasible solution to the MSP model in a finite number of steps. In addition, the proposed strategies are characterized by sleep periods that are up to 40\% greater than those provided by the considered Uniform Power Allocation (UPA) strategy.

\appendix[Regression Model for the Rate Approximation] 
Let $\Hat{r}_1, \ldots, \Hat{r}_T$ be the actual rate values associated to the reception of a TB stream recovered by a user, for different SINR values $\Hat{\sigma}_1, \ldots, \Hat{\sigma}_T$, respectively. By means of a Least Absolute Deviation (LAD) regression model we derive the values of $\alpha$ and $\beta$.

The LTE-A standard imposes to switch to a different MCS as soon as the TB error rate experienced by a user is greater than $10$\%. Hence, for the regression model, we consider only those values of $\Hat{\sigma}_i$ (for $i = 1, \ldots, T$) such that the TB error rate is less than or equal to $0.1$. The LAD model can be expressed as follows~\cite{bloomfield1983least}:
\begin{align}
	\text{(LAD)}&    \quad\quad \mathop{\mathop{\min_{\alpha,\beta}}} \,\, \sum_{i = 1}^{T} \Big| \hat{r}_i - r(\Hat{\sigma}_i) \Big| \label{eq.regression}\\
    \text{subject to}&  \quad\quad \Hat{r}_i \leq r(\Hat{\sigma}_i), \quad\quad\quad\quad\,\,\,\, \text{$i = 1, \ldots, T$} \label{eq.regression.c1} \\
    &  \quad\quad \alpha > 0, \quad \beta > 0  \label{eq.regression.c3} 
\end{align}
where the objective function~\eqref{eq.regression} minimizes the sum of the absolute value of the differences between $\hat{r}_i$ and the value of the fitting function $r(\Hat{\sigma}_i)$. Constraint~\eqref{eq.regression.c1} ensures that the function $r(\sigma)$ does not exceed the simulated TB reception rate.

We define $\sigma_{\emph{min}}$ to be the minimum value of SINR below that a user cannot recover the data flow transmitted with the smallest MCS. Conversely, a user that experiences a SINR equal to or greater than $\sigma_{\emph{max}}$ can always recover a data flow transmitted with the highest MCS. By means of computer simulations we set $\sigma_{\emph{min}}\!=\!6.33 \, \text{dB}$ and $\sigma_{\emph{max}}\! =\! 31.32 \, \text{dB}$~\cite{Access2013}. Then, the correcting factors, obtained by means of the Least Absolute Deviation (LAD) regression model, are $\alpha = 0.17$ and $\beta = 0.06$.

\begin{figure}[tbd] 
		\vspace{-3mm}
       \centering
       \begin{tikzpicture}
			  \node[anchor=south west,inner sep=0] (image) at (0,0) {\includegraphics[width=0.85\columnwidth]{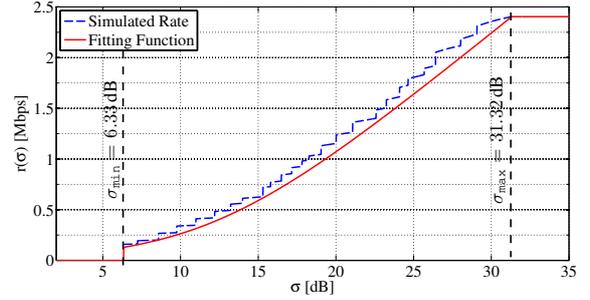}};
 \begin{scope}[x={(image.south east)},y={(image.north west)}]
		  \draw[dashed,semithick] (0.202,0.78) -- (0.202,0.1);
		  \node [draw=none,rotate=90,scale=0.7] at (0.182,0.50) {$\sigma_{\texttt{min}} = $ \textbf{\SI{6.33}{\dB}}};
		
		  \draw[dashed,semithick] (0.878,0.90) -- (0.878,0.13);
		  \node [draw=none,rotate=90,scale=0.7] at (0.855,0.50) {$\sigma_{\texttt{max}} = $ \textbf{\SI{31.32}{\dB}}};
		   \end{scope}
			\end{tikzpicture}    
\caption{eMBMS user reception rate derived by computer simulations and by~\eqref{eq_shannon} vs. $\sigma$.} 
        \label{fig.reg}
        \vspace{0mm}
\end{figure}

Fig.~\ref{fig.reg} compares both the simulated eMBMS user reception rate with the rate values obtained from~\eqref{eq_shannon}, as a function of $\sigma$. That figure clearly shows that the Shannon's rate expression, resulting from the LAD regression model, is a good approximation of the actual eMBMS rate values.

\bibliographystyle{IEEEtran}
\bibliography{IEEEabrv,bib}

\begin{thebibliography}{10}
\providecommand{\url}[1]{#1}
\csname url@samestyle\endcsname
\providecommand{\newblock}{\relax}
\providecommand{\bibinfo}[2]{#2}
\providecommand{\BIBentrySTDinterwordspacing}{\spaceskip=0pt\relax}
\providecommand{\BIBentryALTinterwordstretchfactor}{4}
\providecommand{\BIBentryALTinterwordspacing}{\spaceskip=\fontdimen2\font plus
\BIBentryALTinterwordstretchfactor\fontdimen3\font minus
  \fontdimen4\font\relax}
\providecommand{\BIBforeignlanguage}[2]{{%
\expandafter\ifx\csname l@#1\endcsname\relax
\typeout{** WARNING: IEEEtran.bst: No hyphenation pattern has been}%
\typeout{** loaded for the language `#1'. Using the pattern for}%
\typeout{** the default language instead.}%
\else
\language=\csname l@#1\endcsname
\fi
#2}}
\providecommand{\BIBdecl}{\relax}
\BIBdecl

\bibitem{sesia}
S.~Sesia, M.~Baker, and I.~Toufik, \emph{{LTE - The UMTS Long Term Evolution:
  From Theory to Practice}}.\hskip 1em plus 0.5em minus 0.4em\relax John Wiley
  \& Sons, 2011.

\bibitem{6461191}
M.~Gupta, S.~Jha, A.~Koc, and R.~Vannithamby, ``{Energy Impact of Emerging
  Mobile Internet Applications on LTE Networks: Issues and Solutions},''
  \emph{IEEE Commun. Mag.}, vol.~51, no.~2, pp. 90--97, 2013.

\bibitem{6151867}
S.~Jin and D.~Qiao, ``{Numerical Analysis of the Power Saving in 3GPP LTE
  Advanced Wireless Networks},'' \emph{IEEE Trans. Veh. Technol.}, vol.~61,
  no.~4, pp. 1779--1785, 2012.

\bibitem{5373741}
G.~S. Kim, Y.~H. Je, and S.~Kim, ``{An Adjustable Power Management for Optimal
  Power Saving in LTE Terminal Baseband Modem},'' \emph{IEEE Trans. Consum.
  Electron.}, vol.~55, no.~4, pp. 1847--1853, 2009.

\bibitem{6472116}
J.-M. Liang, J.-J. Chen, H.-H. Cheng, and Y.-C. Tseng, ``{An Energy-Efficient
  Sleep Scheduling With QoS Consideration in 3GPP LTE-Advanced Networks for
  Internet of Things},'' \emph{Emerging and Selected Topics in Circuits and
  Systems, IEEE Journal on}, vol.~3, no.~1, pp. 13--22, 2013.

\bibitem{6154772}
C.~Wu, X.~Sun, and T.~Zhang, ``{A Power-Saving Scheduling Algorithm for Mixed
  Multicast and Unicast Traffic in MBSFN},'' in \emph{Computing, Communications
  and Applications Conference (ComComAp), 2012}, 2012, pp. 170--174.

\bibitem{6025326}
P.~Seeling and M.~Reisslein, ``Video transport evaluation with h.264 video
  traces,'' \emph{Communications Surveys Tutorials, IEEE}, vol.~14, no.~4, pp.
  1142--1165, Fourth 2012.

\bibitem{jsacTassi}
A.~Tassi, I.~Chatzigeorgiou, and D.~Vukobratovi\'c, ``{Resource Allocation
  Frameworks for Network-coded Layered Multimedia Multicast Services},''
  \emph{IEEE J. Sel. Areas Commun.}, 2014.

\bibitem{6353684}
D.~Lecompte and F.~Gabin, ``{Evolved Multimedia Broadcast/Multicast Service
  (eMBMS) in LTE-Advanced: Overview and Rel-11 Enhancements},'' \emph{IEEE
  Commun. Mag.}, vol.~50, no.~11, pp. 68--74, 2012.

\bibitem{6416071}
E.~Magli, M.~Wang, P.~Frossard, and A.~Markopoulou, ``{Network Coding Meets
  Multimedia: A Review},'' \emph{IEEE Trans. Multimedia}, vol.~15, no.~5, pp.
  1195--1212, 2013.

\bibitem{6353397}
C.~Khirallah, D.~Vukobratovi\'c, and J.~Thompson, ``{Performance Analysis and
  Energy Efficiency of Random Network Coding in LTE-Advanced},'' \emph{IEEE
  Trans. Wireless Commun.}, vol.~11, no.~12, pp. 4275--4285, 2012.

\bibitem{fragouli}
C.~Fragouli, ``{Network Coding: Beyond Throughput Benefits},'' \emph{Special
  Issue on Network Coding at the Proceedings of the IEEE}, 2011.

\bibitem{5634159}
O.~Trullols-Cruces, J.~Barcelo-Ordinas, and M.~Fiore, ``{Exact Decoding
  Probability Under Random Linear Network Coding},'' \emph{IEEE Commun. Lett.},
  vol.~15, no.~1, pp. 67--69, Jan. 2011.

\bibitem{6550868}
F.~Chiti, R.~Fantacci, F.~Schoen, and A.~Tassi, ``{Optimized Random Network
  Coding for Reliable Multicast Communications},'' \emph{Communications
  Letters, IEEE}, vol.~17, no.~8, pp. 1624--1627, 2013.

\bibitem{Rel9}
{3GPP TR 36.814 v9.0.0 (Release 9)}, ``{Further Advancements for (E-UTRA)},''
  2010.

\bibitem{Munaretto}
D.~Munaretto, D.~Jurca, and J.~Widmer, ``{Broadcast Video Streaming in Cellular
  Networks: An Adaptation Framework for Channel, Video and AL-FEC Rates
  Allocation},'' in \emph{Proc. of WICON 2010}, Singapore, CN, Mar. 2010, pp.
  1--9.

\bibitem{bloomfield1983least}
P.~Bloomfield and W.~L. Steiger, \emph{{Least Absolute Deviations: Theory,
  Applications, and Algorithms}}.\hskip 1em plus 0.5em minus 0.4em\relax
  Birkh{\"a}user Boston, 1983.

\bibitem{Access2013}
S.~Schwarz, J.~Ikuno, M.~Simko, M.~Taranetz, Q.~Wang, and M.~Rupp, ``Pushing
  the limits of {LTE}: A survey on research enhancing the standard,''
  \emph{{IEEE} Access}, vol.~1, pp. 51--62, 2013.

\end{thebibliography}
\end{document}